\documentclass[final,1p,times]{elsarticle}

\usepackage{graphicx}
\usepackage{amssymb}

\begin{document}

\begin{frontmatter}


\title{Wavepacket approach to particle diffraction by thin
targets: \\ quantum trajectories and arrival times}
\author[ceft]{C. Efthymiopoulos}
\ead{cefthim@academyofathens.gr}
\author[ndel]{N. Delis}
\ead{delnike@gmail.com}
\author[gcon]{G. Contopoulos}
\ead{gcontop@academyofathens.gr}
\address[ceft,ndel,gcon]{Research Center for Astronomy and
Applied Mathematics, Academy of Athens}
\address[ndel]{Department of Physics, University of Athens,
Panepistimiopolis, 153 42 Athens, Greece}

\begin{abstract}
We develop a wavepacket approach to the diffraction of charged
particles by a thin material target and we use the de Broglie-Bohm
quantum trajectories to study various phenomena in this context.
We construct a particle wave function model given as the sum
of two terms $\psi=\psi_{ingoing}+\psi_{outgoing}$, each having
a wavepacket form with longitudinal and transverse quantum
coherence lengths both finite. We find the form of the separator,
i.e.the limit between the domains of prevalence of the ingoing
and outgoing quantum flow. The structure of the quantum-mechanical
currents in the neighborhood of the separator implies the formation
of an array of \emph{quantum vortices} (nodal point - X point
complexes). The X point gives rise to stable and unstable manifolds,
whose directions determine the scattering of the de Broglie - Bohm
trajectories. We show how the deformation of the separatior near Bragg
angles explains the emergence of a diffraction pattern by the de
Broglie - Bohm trajectories. We calculate the arrival time distributions
for particles scattered at different angles. A main prediction is that
the arrival time distributions have a dispersion proportional to
$v_0^{-1}\times$ the largest of the longitudinal and transverse
coherence lengths, where $v_0$ is the mean velocity of incident
particles. We also calculate time-of-flight differences $\Delta T$
for particles scattered in different angles. The predictions of
the de Broglie - Bohm theory for $\Delta T$ turn to be different
from estimates of the same quantity using other theories on time
observables like the sum-over-histories or the Kijowski approach.
We propose an experimental setup aiming to test such predictions.
Finally, we explore the semiclassical limit of short wavelength
and short quantum coherence lengths, and demonstrate how, in this
case, results with the de Broglie - Bohm trajectories are similar
to the classical results of Rutherford scattering.
\end{abstract}


\end{frontmatter}

\section{Introduction}

The {\it de Broglie - Bohm quantum  trajectories} \cite{debro1928}
\cite{bohm1952}\cite{bohmhil1993}\cite{hol1993} have been
considered as an interpretational tool in a number of recent applications
(see \cite{wya2005}\cite{durteu2009}\cite{cha2010} for reviews),
since they can offer new insight into a variety of complex quantum
phenomena. According to the de Broglie-Bohm theory, to any
wavefunction $\psi(\mathbf{r}_1,\mathbf{r}_2,\ldots,\mathbf{r}_N,t)$
describing a $N-$particle system, we can associate a set of
`quantum trajectories'. One trajectory is defined by the initial
conditions $(\mathbf{r}_1(0),\mathbf{r}_2(0),\ldots,\mathbf{r}_N(0))$
and by the `pilot wave' equations of motion
\begin{equation}\label{sch2}
{d\mathbf{r}_i\over dt}={\hbar\over
m_i}Im({\nabla_i\psi\over \psi}),~~~i=1,\ldots N
\end{equation}
where $m_i$ are the particle masses and $\hbar$ is Planck's
constant. The equations of motion (\ref{sch2}) imply the
continuity equation for the probability density
$\rho(\mathbf{r}_1,\mathbf{r}_2,\ldots,\mathbf{r}_N,t) =
|\psi(\mathbf{r}_1,\mathbf{r}_2,\ldots,\mathbf{r}_N,t)|^2$.
In particular, in a one-particle system we can choose many
different initial conditions corresponding to an initial
density $\rho(\mathbf{x},0)=|\psi(\mathbf{r},0)|^2$. Then,
the pilot-wave equations guarantee the preservation of Born's
rule ${\rho(\mathbf{r},t)}=|\psi(\mathbf{r},t)|^2$ at all
subsequent times $t$. Furthermore, the de Broglie - Bohm
trajectories are equivalent to the stream lines of the
quantum probability current $\mathbf{j}=(\hbar/2mi)
(\psi^*\nabla\psi-\psi\nabla\psi^*)$. Thus, the Bohmian
approach yields practically equivalent results to Madelung's
quantum hydrodynamics \cite{mad1926}.

The de Broglie - Bohm theory has been discussed extensively
from the point of view of its relevance as a consistent
interpretation of quantum mechanics (e.g.
\cite{hol1993}, \cite{bacval2009}; see \cite{tow2011} for an
extended list of references). However, the employment of
the de Broglie - Bohm trajectories has been proven useful also
in many {\it practical} aspects of the study of quantum systems.
Some modern applications are:

i) Visualization of quantum processes: examples are barrier
penetration or the quantum tunneling effect \cite{hiretal1974a}
\cite{dewhil1982}\cite{skoetal1989}\cite{lopwya1999},
the (particle) two-slit experiment \cite{phietal1979}, ballistic
transport through `quantum wires' \cite{beehou1991}\cite{beretal2001},
molecular dynamics \cite{gin2003}, dynamics in nonlinear systems
with classical focal points or caustics \cite{zhamak2003}, and
rotational or atom-surface scattering \cite{ginetal2002}
\cite{sanzetal2004a}\cite{sanzetal2004b}.

ii) Lagrangian solvers of Schr\"{o}dinger equation via swarms
of evolving Bohmian trajectories (see \cite{wya2005} for a
comprehensive review, as well as \cite{sanzetal2002}
\cite{sanzetal2004a}\cite{sanzetal2004b}\cite{ori2007}).
The interest in this method lies in that, instead of solving
Schr\"{o}dinger's equation first, one uses a step-by-step procedure
to calculate the trajectories via Newton's second order equations
of motion in a potential
\begin{equation}\label{sch3}
{U(\mathbf{r},t)}=V(\mathbf{r},t)+Q(\mathbf{r},t)
\end{equation}
where ${Q(\mathbf{r},t)} $ is the `quantum potential',
caused by the wavefunction $\psi$:
\begin{equation}\label{sch4}
{Q(\mathbf{r},t)}=-{\hbar^2\over 2m}{\nabla^2|\psi|\over |\psi| }.
\end{equation}
Using the information of the initial value of the wavefunction
as well as the evolution of the quantum trajectories, the
wavefunction can then be determined at any subsequent time
step.

iii) Dynamical origin of the {\it quantum relaxation} \cite{valwes2005}
\cite{eftcon2006}\cite{ben2010}\cite{colstru2010}\cite{towetal2011}.
The de Broglie - Bohm theory offers a justification of Born's rule
$\rho=|\psi|^2$, since it predicts that, under some conditions,
the quantum trajectories lead to an asymptotic (in time) approach
towards this rule even if it was initially allowed that
$\rho_{initial}\neq |\psi_{initial}|^2$. It should be
noted that not all choices of $\rho_{initial}$ are guaranteed
to lead to quantum relaxation, and counter-examples can be found,
for reasons explained in \cite{eftcon2006}. The arguments used in
that paper to explain the suppression of the quantum relaxation effect
in the two-slit experiment apply also to many other cases (see e.g.
\cite{sanzetal2000}). In particular, a necessary condition for
quantum relaxation to take place is that the trajectories
should exhibit {\it chaotic} behavior (see \cite{valwes2005}
\cite{eftcon2006}); however, even this condition is not sufficient
(see \cite{conetal2011}). The problem of chaos in the de Broglie - Bohm
theory has been studied extensively (indicative references are
\cite{duretal1992}\cite{faisch1995}\cite{parval1995}\cite{depol1996}
\cite{dewmal1996}\cite{iacpet1996}\cite{fri1997}\cite{konmak1998}
\cite{wuspru1999}\cite{maketal2000}\cite{cush2000}\cite{desalflo2003}
\cite{falfon2003}\cite{wispuj2005}\cite{wisetal2007}\cite{schfor2008}).
We have worked on this problem in \cite{eftcon2006}\cite{eftetal2007}
\cite{coneft2008}\cite{eftetal2009}\cite{deletal2011}.
Our main result was that chaos is due to the presence of
{\it moving quantum vortices} forming `nodal point - X-point
complexes' (\cite{eftetal2007}\cite{coneft2008}\cite{eftetal2009};
see also \cite{wispuj2005}\cite{wisetal2007}). Quantitative
studies of chaos and of the effects of vortices are presented in
\cite{fri1997}\cite{schfor2008}\cite{eftetal2009}. In particular,
in \cite{eftetal2009} we made a theoretical analysis of the
dependence of Lyapunov exponents of the quantum trajectories on
the size and speed of the quantum vortices, thus explaining
numerical results found in \cite{eftetal2007} and \cite{coneft2008}.
Furthermore, in \cite{eftcon2006} we gave examples of systems
which do or do not exhibit quantum relaxation, depending on
whether or not their underlying trajectories are chaotic.
It should be emphasized that, besides chaos, the quantum
vortices play a key role in a variety  of quantum dynamical
phenomena (e.g. \cite{mccwya1971}\cite{hiretal1974a}
\cite{hiretal1974b}\cite{sanzetal2004a}\cite{sanzetal2004b}).

iv) {\it Arrival times and times of flight}. In the traditional
formulation of quantum mechanics time is only a parameter
in Schr\"{o}dinger's equation, since by a theorem of Pauli
\cite{pau1926} no definition of a self-adjoint time-operator
consistent with all axioms of quantum mechanics can be given
in a system with energy spectrum bounded from below. Time,
however, is an experimental observable. Various approaches in
the literature, reviewed in \cite{muglea2000}\cite{mugetal2002},
have addressed the question of a consistent definition of quantum
probability distributions for time observables. Besides the
Bohmian approach, two other approaches are: a) the
`sum-over-histories' approach \cite{hartle1988}\cite{yamtak1993}
based on Feynman paths, and b) the approach of Kijowski
\cite{kij1974}, based on the definition of quantum states acted
upon by the so-called 'Bohm-Aharonov operator' (see
\cite{muglea2000}). On the other hand, the de Broglie - Bohm
approach gives a straightforward answer to this problem,
since the time needed to connect any two points along a quantum
trajectory is a well defined quantity (see \cite{lea1990a}
\cite{lea1990b}).

Regarding this latter point, a key remark that will concern us
in the sequel is that a consistent definition of the arrival times,
that would allow in principle for a comparison of the various
approaches in specific quantum systems, is only possible provided
that the initial wavefunction is {\it localized in space}, i.e. it
is described by a wavepacket model.

Being motivated by the latter remark, in the present paper we present
a study of the de Broglie - Bohm trajectories in a wavepacket model
referring to a quantum phenomenon that has played a fundamental role
in the development of quantum mechanics, namely the {\it diffraction}
of charged particles (e.g. electrons or ions) by a thin material target.

A theoretical study on the quantum scattering problem in the framework
of the de Broglie - Bohm approach has been presented in the series of
works \cite{dau1996}\cite{dauetal1996}\cite{dauetal1997}
\cite{durretal2000}\cite{durretal2006}.
These studies refer to the establishment of the rules of scattering
probabilities using the `flux across surfaces' theorem adapted
to the concept of quantum trajectories. However, they do not deal
with the form of the quantum trajectories or the emergence of diffraction
patterns under specific scattering potentials. A numerical simulation
of Rutherford scattering by a single nucleus has been presented in
\cite{mei2006}, while in \cite{sanzetal2002} the phenomena of
atom-surface scattering as well as neutron diffraction by
slits are considered, which share some common features, but also
important differences, with our problem.

In the present paper we make a detailed study of the de Broglie
- Bohm trajectories in the context of a wavepacket model of
charged particle diffraction, by first investigating the form
of the {\it quantum currents} corresponding to various cases
of this model. These cases are diversified one from the other
by the different quantitative relations characterizing the so-called
{\it quantum coherence lengths} in the longitudinal and transverse
directions of the charged particle beam. This is necessary in order
to be able to compare the results corresponding to possibly different
experimental realizations of a charged particle beam, as e.g. in the
case of electrons produced either by thermionic or by a cold-field
emission processes.

Our present study completes in a substantial way the study initiated
in a previous paper of ours  \cite{deletal2011}, in which we implemented
the de Broglie - Bohm approach in the case of electron diffraction through
a thin crystal. In that study, however, we assumed a planar wave model
for the propagation of the electron wavefunction in the longitudinal
direction. In contrast, in the present paper we assume instead
a finite longitudinal quantum coherence length. This assumption leads to
a number of crucial new elements with respect to \cite{deletal2011}.
In fact, in order to achieve our goal we derive a wavefunction model
by a refined implementation of basic scattering theory, so as to
account for a fully-localized in space description of scattering.
The derivation of this model presents its own interest, and it
is exposed in detail in section 2.

\begin{figure}
\centering
\includegraphics[scale=0.65]{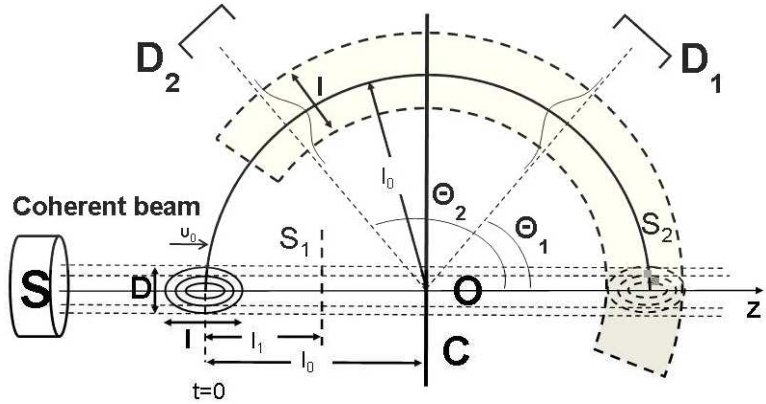}
\caption{The basic setup of the problem under study. A source
(S) emits charged particles described by an `ingoing' wavefunction
having the form of a wavepacket with dispersions $l$ in the
longitudinal direction (z-axis $\equiv$ direction of incidence to
a thin material target (C) placed at the center O of the coordinate
system), and $D$ in the transverse direction. After scattering,
some particles arrive at detectors $D_i$ placed at equal
distances from O and various angles $\theta_i$. The wavefunction
is assumed to have axial symmetry (around the z-axis), thus
the figure corresponds to any meridian plane. Various other
symbols are explained in the text.}
\label{setup}
\end{figure}
The structure of the paper is as follows: after the derivation of
the basic wavefunction model in section 2, we pass to a study of the
quantum trajectories in section 3. Here the emphasis is on the
influence upon the trajectories of quantum vortices, whose
appearance and role in this problem are explicitly discussed.
In fact, we show that the quantum vortices appear in the transition
zone from a domain of predominance of the ingoing wavefunction to
a domain of predominance of the outgoing wavefunction. Inside this
zone we can define a locus called {\it separator}, which plays
a key role in the interpretation of the scattering process via
the quantum trajectories. In section 4 we study the arrival times of
diffracted particles to detectors placed in various scattering angles.
A main outcome of this study is that it is possible to propose a
feasible experimental test probing the predictions of the Bohmian theory
about the particles' arrival times. In section 5 we discuss separately
the `semi-classical' case of particles with a large mass and and a
small de Broglie wavelength, applicable e.g. to $\alpha-$particle
or ion scattering, since this case exhibits some special features
in comparison to the case of electron diffraction. Finally,
section 6 summarizes the main conclusions of the present study.

\section{Modelling of the wavefunction}
We consider a cylindrical beam of particles of mass $m$ and
charge $Z_1q_e$ incident on a thin material target. We set the
center of the target as the origin of our coordinate system
of reference, and use both cylindrical coordinates
$(z,R,\phi)$ and spherical coordinates $(r,\theta,\phi)$.
The $z-$axis is the beam's main axis, $R$ denotes cylindrical
radius transversally to $z$, $\phi$ is the azimuth,
$r=(z^2+R^2)^{1/2}$ and $\theta=\tan^{-1}(R/z)$
(see Figure 1, schematic).

A basic form of diffraction theory for charged particles, reviewed
e.g. in \cite{peng2005}, assumes that the incident waves are planar.
As explained in the introduction, here instead we are interested
in a wavepacket approach. Focusing only on elastic scattering
phenomena, the latter approach can be obtained by a refinement
of the basic theory as follows:

The potential felt by a charged particle approaching the
target can be considered as the sum of the individual potential
terms generated by every atom in the target:
\begin{equation}\label{pot}
V(\mathbf{r})=\sum_{j=1}^N U(\mathbf{r}-\mathbf{r}_j)~~.
\end{equation}
where $\mathbf{r}_j$ denotes the position of j-th atom in the
lattice of the target (this position exhibits some statistical
fluctuations due to thermal oscillations etc; the effect of
these fluctuations is discussed later in this section). As a
model for the function $U$, we can adopt a screened Coulomb
potential
\begin{equation}\label{potatom}
U(\mathbf{r-r_j})={1\over 4\pi\epsilon_0}
{Z_1Zq_e^2\exp(-|\mathbf{r-r_j}|/r_0)\over|\mathbf{r-r_j}|}
\end{equation}
($\epsilon_0$ = vacuum dielectric constant), where $Z$ is the
nuclear charge, and $r_0$ is a constant representing a charge
screening range within the atoms, whose value is of the order
of the atomic size.\\

Particles being scattered by the target can be described by
a wavefunction given as a superposition of eigenfunctions
\begin{equation}\label{psiall}
\psi(\mathbf{r},t)={1\over (2\pi)^{3/2}} \int
d^3\mathbf{k}~\tilde{c}(\mathbf{k})\phi_{\mathbf{k}}
(\mathbf{r})e^{{-i\hbar k^2t/2m}}
\end{equation}
where $\tilde{c}(\mathbf{k})$ are Fourier coefficients, and
$\phi_{\mathbf{k}}(\mathbf{r})$ are scattering eigenfunctions,
i.e. solutions of the time-independent Schr\"{o}dinger's equation
\begin{equation}\label{sch1}
-{\hbar^2\over 2m}\nabla^2 \phi + V(\mathbf{r})\phi = E\phi
\end{equation}
with $V$ chosen as in (\ref{pot}) and $E>0$. The different
solutions $\phi\equiv\phi_\mathbf{k}$ are labeled by their wavevectors
$\mathbf{k}$ of modulus $k\equiv \mid \mathbf{k}\mid=(2mE)^{1/2}/\hbar$,
where $E>0$ is the energy associated with one eigenstate.
Born's approximation can be used to obtain an approximative
formula for $\phi_{\mathbf{k}}$. We thus write
\begin{equation}\label{bornser}
\phi_\mathbf{k}=
\phi_{0,\mathbf{k}}+\phi_{1,\mathbf{k}}+\phi_{2,\mathbf{k}}
+\ldots
\end{equation}
where $\phi_{0,\mathbf{k}}=e^{i\mathbf{k r}}=O(1)$ is the solution
of Eq.(\ref{sch1}) for the free particle problem $(
V(\mathbf{r})=0)$, while $\phi_{1,\mathbf{k}}=O(V)$,
$\phi_{2,\mathbf{k}}=O(V^2)$ etc (assuming that $V$ small compared
to the particles' energies). The above series are meaningful at all
points of space excluding a set of balls of radius a few times $r_0$
around every one of the atoms in the target. Spherical harmonic
expansions (see e.g. \cite{mes1961}) provide a more accurate
representation of the solution inside such balls, but their use
is cumbersome while practically unnecessary in the context of
the present study.

A step by step determination of the series terms in (\ref{bornser})
can be obtained via the recursive formula
\begin{equation}\label{sch1ret}
-{\hbar^2\over 2m}\nabla^2 \phi_{n,\mathbf{k}} +
V\phi_{n-1,\mathbf{k}}= E\phi_{n,\mathbf{k}}={\hbar^2k^2\over
2m}\phi_{n,\mathbf{k}}~~.
\end{equation}
All essential phenomena discussed below are present already in the
solutions including just the two first terms
$\phi_{\mathbf{k}}\simeq \phi_{0,\mathbf{k}}+\phi_{1,\mathbf{k}}$.
From Eq.(\ref{sch1ret}) for $n=1$ we find:
\begin{equation}\label{phi1}
\phi_{1,\mathbf{k}}(\mathbf{r})=-{m\over 2\pi\hbar^2}
\int_{\mbox{all space}} d^3\mathbf{r'}
{e^{ik|\mathbf{r-r'}|}\over|\mathbf{r-r'}|} \left(e^{i\mathbf{k\cdot
r'}}\sum_{j=1}^N{1\over 4\pi\epsilon_0}
{Z_1Zq_e^2e^{-|\mathbf{r'-r_j}|/r_0}\over|\mathbf{r'-r_j}|} \right)~~.
\end{equation}
The integral in (\ref{phi1}) can be estimated using standard
approximations of scattering theory. We then find
\begin{equation}\label{phiall}
\phi_{\mathbf{k}}(\mathbf{r})\simeq e^{i\mathbf{k\cdot r}}
-{Z_1Zq_e^2\over 4\pi\epsilon_0}{m\over \hbar^2}
\left(\sum_{j=1}^N{e^{ik\mid\mathbf{r-r_j\mid}}e^{i\mathbf{k\cdot
r_j}} \over
\mid\mathbf{r-r_j\mid}(2k^2\sin^2(\Delta\theta_j/2)+1/2r_0^2)}\right)
\end{equation}
where $\Delta\theta_j$ denotes the angle between the vectors $\mathbf{k}$
and $\mathbf{r}-\mathbf{r_j}$.

Substituting Eq.(\ref{phiall}) into Eq.(\ref{psiall}) we have
\begin{eqnarray}\label{psiall2}
\psi(\mathbf{r},t)&\simeq&{1\over (2\pi)^{3/2}}\Bigg\{ \int
d^3\mathbf{k}~\tilde{c}(\mathbf{k})e^{i\mathbf{k
r}}e^{{-i\hbar k^2t/2m}}\\
&-&{Z_1Zq_e^2\over4\pi\epsilon_0}{m\over \hbar^2}
\int d^3\mathbf{k}~\tilde{c}(\mathbf{k})
\left(\sum_{j=1}^N{e^{ik\mid\mathbf{r-r_j\mid}}e^{i\mathbf{k\cdot
r_j}} \over
\mid\mathbf{r-r_j\mid}(2k^2\sin^2(\Delta\theta_j/2)+1/2r_0^2)}\right)
e^{{-i\hbar k^2t/2m}}\Bigg\}\nonumber
\end{eqnarray}
The problem of defining $\psi(\mathbf{r},t)$ is now restricted to
making an appropriate choice for the coefficients $\tilde{c}(\mathbf{k})$.
The latter are determined by the Fourier transform of the initial
wavefunction $\psi(\mathbf{r},t=0)$. In the wavepacket approach,
the initial wavefunction is localized around the source, i.e. far
from the target. Hence we can set
$\psi(\mathbf{r},t=0)\simeq\psi_{ingoing}(\mathbf{r},t=0)$, where
$\psi_{ingoing}(\mathbf{r},t=0)$ represents a wavepacket moving in
the z-direction towards the target with some velocity $v_0$.
A Gaussian wavepacket of this form corresponds (in momentum space)
to the choice
\begin{equation}\label{psimom}
\tilde{c}(\mathbf{k})={1\over \pi^{1/2}\sigma_\perp}
{1\over \pi^{1/4}\sigma_\parallel^{1/2}}
\exp\left(-{k_x^2+k_y^2\over 2\sigma_\perp^2})\right)
\exp\left(-{(k_z-k_0)^2\over 2\sigma_\parallel^2}-ik_zz_0\right)~~.
\end{equation}
In (\ref{psimom}), $(k_x,k_y,k_z)$ are the Cartesian components of
$\mathbf{k}$, $z_0=-l_0$ is the initial position of the center of the
wavepacket along the z-axis, and $k_0=m v_0/\hbar$. The quantities
$\sigma_\parallel$, $\sigma_\perp$ are the longitudinal and transverse
dispersions of the wavepacket in momentum space. These correspond
to dispersions in position space given by $l=\sigma_\parallel^{-1}$
and $D=\sigma_\perp^{-1}$. The quantities $l$ and $D$ are hereafter
called the longitudinal and transverse quantum coherence length
respectively. Eq.(\ref{psiall2}) now takes the form
\begin{equation}\label{psiinout}
\psi(\mathbf{r},t)=\psi_{ingoing}(\mathbf{r},t)
+\psi_{outgoing}(\mathbf{r},t)
\end{equation}
where
\begin{equation}\label{psiin}
\psi_{ingoing}= B(t)
\exp\left(-{R^2\over 2(D^2+{i\hbar t\over m})}
-{(z+l_0-{\hbar k_0\over m}t)^2\over 2(l^2+{i\hbar t\over m})}+ik_0z\right)
\end{equation}
with
$$
B(t)={1\over \pi^{3/4}}\left({D\over D^2+i\hbar t/m}\right)
\left({l\over l^2+i\hbar t/m}\right)^{1/2}
\exp\left(ik_0l_0-{i\hbar k_0^2\over 2m}t\right)
~~.
$$
The function $\psi_{outgoing}$ corresponds to the second integral
in (\ref{psiall2}). An explicit expression for this function can
only be found by adopting some further approximations. First,
we consider fast-moving wavepackets, for which $k_0\gg
\max(\sigma_\perp,\sigma_\parallel)$ as well as $k_0\gg 1/r_0$.
Then, in the denominator of the second integrand in (\ref{psiall2}):
i) the term $1/2r_0^2$ can be ignored, and ii) we use
the approximation $1/k^2\simeq 1/k_0^2$. Second, at all distances
$r\gg r_j$ we have that the angles $\Delta\theta_j$ are
approximately equal one to the other and to the angle
$\theta$ (which is equal to the angle between the vectors
$\mathbf{r}$ and $\mathbf{k}_0=(0,0,k_0)$. Finally, we set
$\mid\mathbf{r-r_j}\mid\approx r-\mathbf{r_j\cdot n}+
r_j^2/(2r)$ in the exponential argument of (\ref{psiall2}),
where $\mathbf{n}=(\sin\theta\cos\phi,\sin\theta\sin\phi,\cos\theta)$
(this is necessary in order to retain all terms whose phase
has a substantially non-zero value), while we set
$\mid\mathbf{r-r_j}\mid\approx r$ in the denominator of
the integrands in (\ref{psiall2}). Using these approximations,
we find
\begin{eqnarray}\label{psiout}
\psi_{outgoing}&\approx& B(t){Z_1Zq_e^2\over 4\pi\epsilon_0}
{m\over \hbar^2}{1\over 2k_0^2\sin^2(\theta/2)}{\exp(ik_0r)\over r}
\nonumber\\
&\times&\sum_{j=1}^N\Bigg[
\exp\left(ik_0[-\mathbf{r_j\cdot n}+z_j+r_j^2/(2r)]\right)\\
&~&~~~~~~~~\exp\left(-{R_j^2\over 2(D^2+{i\hbar t\over m})}
-{(r+l_0-v_0t-\mathbf{r_j\cdot n}+z_j+{r_j^2\over 2r})^2
\over 2(l^2+{i\hbar t\over m})}\right)\Bigg]\nonumber
\end{eqnarray}
where $v_0=\hbar k_0/m$ represents the mean velocity of a particle
with wavenumber $k_0$.

At distances $r$ closer to the target than the maximum of the two
coherence lengths $D,l$, the prefactor $f(r,\theta)=1/(2k_0^2\sin^2(\theta/2)r)$
in Eq.(\ref{psiout}) is no longer accurate. In order to be able to
perform some numerical calculations of de Broglie - Bohm trajectories,
after numerically simulating the sums appearing in (\ref{psiall2}) we
found by trial a fitting model that represents reasonably well the
modifications of $f(r,\theta)$ close to the target. This reads:
\begin{equation}\label{fit}
f(r,\theta)=k_0^{-2}\left[c_3D\sin\theta+(c_3^2D^2\sin^2\theta
+r^2-2rc_4D\sin\theta+c_4^2D^2)^{1/2}-r\cos\theta\right]^{-1}
\end{equation}
where $c_3$ and $c_4$ are fitting constants determined by comparison of
Eq.(\ref{fit}) to the results of the numerical simulation of $f(r,\theta)$
(in all simulations below we set $c_3=0.3$, $c_4=0.8$).
It is to be stressed that Eq.(\ref{fit}) correctly recovers the asymptotic
form $f\sim 1/(2k_0^2\sin^2(\theta/2)r)$ when $r$ is large.

The outgoing wavefunction now takes the form
\begin{equation}\label{psioutl2}
\psi_{outgoing}\approx \ {B(t)Z_1Zq_e^2m\over
4\pi\epsilon_0\hbar^2}
e^{ik_0r} f(r,\theta)
S_{eff}(k_0,\mathbf{r},t)
\end{equation}
where the quantity $S_{eff}(k_0;\mathbf{r})$ is called hereafter
the `effective Fraunhoffer function' (in analogy with the
`far field' diffraction limit in wave optics, see \cite{ers2007}).
This is given by
\begin{eqnarray}\label{seff}
S_{eff}(k_0,\mathbf{r},t)&=&
\sum_{j=1}^N\Bigg[
\exp\left(ik_0(-\mathbf{r_j\cdot n}+z_j+r_j^2/(2r))\right)\\
&~&~~~~~~~~\exp\left(-{R_j^2\over 2(D^2+{i\hbar t\over m})}
-{(r+l_0-v_0t-\mathbf{r_j\cdot n}+z_j+{r_j^2\over 2r})^2
\over 2(l^2+{i\hbar t\over m})}\right)\Bigg]~~.\nonumber
\end{eqnarray}
The physical significance of the function $S_{eff}(k_0;\mathbf{r},t)$
is that it sums the contributions of all the atoms in the target which
act as sources of partial outgoing waves, whose superposition forms
$\psi_{outgoing}$. Furthermore, the function $S_{eff}$ accounts for
the formation of a diffraction pattern, which, for given $\mathbf{n}$,
$r$, arises by the coherent contributions of all atoms in the target
whose phasors $\exp[ik_0(-\mathbf{r_j\cdot n}+z_j+r_j^2/(2r))]$
are nearly parallel one to the other. As in \cite{deletal2011}, we
consider the simplest example of a cubic lattice structure of the
target
\begin{eqnarray}\label{lattice}
&~&\mathbf{r}_j=(n_x,n_y,n_z)a + \Delta a\mathbf{u}_j(t), \nonumber\\
&~&\\
(n_x,n_y,n_z)&\in& (-{N_\perp\over 2},{N_\perp\over 2})\times
(-{N_\perp\over 2},{N_\perp\over 2})\times (-{N_z\over 2},{N_z\over
2})\nonumber
\end{eqnarray}
where i) $a$ is the lattice constant (equal to the length of one side
of the primitive cell) ii) $\Delta a$ is the amplitude of some random
oscillations (due to thermal or recoil motions; $\Delta a$ is taken
equal to a small fraction of $a$) and
$\mathbf{u_j}\equiv(u_{j,x},u_{j,y},u_{j,z})$ are random variables
with a uniform distribution in the intervals $[-0.5,0.5]$ (the
random oscillations introduce a so-called Debye-Waller effect,
analyzed in \cite{deletal2011}; here, for simplicity, we ignore
modifications on the wavefunction due to this effect).
iii) The number of atoms $N_z$ in the z-direction is $N_z=d/a$,
where $d$ is the target thickness, and iv) the value of $N_\perp$
is of order $N_\perp=O(D/a)$, due to the Gaussian factor
$\exp(-R_j^2/(2D^2+i\hbar t/m))$ in Eq.(\ref{seff}) which
can be approximated by $\approx 1$  for all $|n_x|<N_\perp/2$
and $|n_y|<N_\perp/2$, and by $0$ for $|n_x|>N_\perp/2$
or $|n_y|>N_\perp/2$ (for typical magnitudes of $D$ the inequality
$D^2>>\hbar t/ m$ holds for all times $t$ of interest in our
study, see below). We now distinguish the following cases:

\subsection{$l>>D>>a$}

When the longitudinal coherence length $l$ is larger than the transverse
coherence length $D$, a simple modeling of the sum in Eq.(\ref{seff})
becomes possible at all distances $r>D$. Ignoring first the random
fluctuations in (\ref{lattice}) (i.e. setting $\Delta a=0$), we have
$r>>|-\mathbf{r_j\cdot n}+z_j+{r_j^2\over 2r}|$ whereby it follows
that
\begin{eqnarray}\label{sefflong}
&~&S_{eff}(k_0,\mathbf{r},t)\approx
\exp\left(-{(r+l_0-v_0t)^2\over 2(l^2+i\hbar t/m)}\right)\nonumber\\
&\times&\sum_{n_x=-N_\perp/2}^{N_\perp/2}
\sum_{n_y=-N_\perp/2}^{N_\perp/2}
\exp\left(ik_0[-an_x\sin\theta\cos\phi-an_y\sin\theta\sin\phi
+{n_x^2a^2+n_y^2a^2)\over 2r}]\right)\\
&\times&\sum_{n_z=-N_z/2}^{N_z/2}
\exp\left(ik_0[(1-\cos\theta)n_za+{n_z^2a^2\over 2r}]\right)\nonumber
\end{eqnarray}
For a random choice of $k_0,\theta,\phi$, the total number of
contributing atoms in the sums of Eq.(\ref{sefflong}) is of the
order $N\sim N_\perp^2N_z=D^2d/a^3$. Furthermore, the $N$ phasors
have an effectively random phase. Thus, the total sum is of order
$N^{1/2}$, and we are lead to the simple estimate
$$
S_{eff}\sim Dd^{1/2}/a^{3/2}\exp[-(r+l_0-v_0 t)^2/(2(l^2+i\hbar t/m))]
$$
called, hereafter, the `diffuse term' of the effective Fraunhofer
function. Using this term we have
\begin{equation}\label{psioutl2}
\psi_{outgoing}\simeq \ {B(t)Z_1Zq_e^2mDd^{1/2}\rho^{3/2}\over
4\pi\epsilon_0\hbar^2}
\exp\left(-{(r+l_0-v_0 t)^2\over{2(l^2+{i\hbar t\over m})}}\right)
f(r,\theta)e^{ik_0r}
\end{equation}
where $\rho=a^{-3}$ is the number density of the atoms in the target.

The model (\ref{psioutl2}) is used in a number of numerical
simulations below. It describes a {\it radial pulse} propagating
outwards with speed $v_0$, which emerges from the center at the
time $l_0/v_0$. It should be stressed, however, that
Eq.(\ref{psioutl2}) requires modifications close to particular
angles where the phasors in the sums of (\ref{sefflong}) are
added coherently. This will be examined in subsection 3.3.
Further modifications are required when $l$ and $D$ become
comparable or smaller than the inter-atomic distance $a$ in
the target. This case is examined in section 5.

\subsection{$D>>l>>a$}

The modeling of $S_{eff}$ is a more subtle problem if the transverse
coherence length is larger than the longitudinal coherence length.
Without loss of generality, we can consider a fixed meridian plane
defined e.g. by the angle $\phi=0$. We set $\xi=r+l_0-v_0t$ and we
make the change of variables $u=-x\sin\theta+(x^2+y^2)/(2r)$,
$R=(x^2+y^2)^{1/2}$. Considering now a random choice of
$k_0,\theta$, and ignoring the quantity $i\hbar t/m$,
for a given value of $\xi$ the sum over the variables $n_x,n_y$
in Eq.(\ref{seff}) can by approximated by a sum over a domain
of values of $(n_x,n_y)$ such that $u(n_x,n_y)$ belongs to a
ball of radius $l$ around $\xi$. The total number of contributing
atoms is then of order $N\sim l^2d/a^3$. If i) we consider the sum
over the phasors as a sum of random numbers (yielding a total
magnitude $\sim N^{1/2}$), and ii) we substitute the second
exponential in (\ref{seff}) by a delta function around $\xi$,
we find $S_{eff}\approx (ld^{1/2}/a^{3/2}) I$, where
\begin{equation}
I=\int\int J(u,R) e^{-R^2/2D^2}\delta(u+\xi)dudR
\end{equation}
and $J(u,R)$ is the determinant of the Jacobian matrix of the
transformation $(x,y)\rightarrow (u,R)$. We find
\begin{equation}\label{iseff}
I=\int_{R_{min}}^{R_{max}}
{e^{-R^2/2D^2}\over \sin\theta
\sqrt{1-{1\over sin^2\theta}\left({R\over 2r}+{\xi\over R}\right)^2}}
dR
\end{equation}
where $R_{min}=|r(\sin\theta-\sqrt{\sin^2\theta-2\xi/r})|$,
$R_{max}=r(\sin\theta+\sqrt{\sin^2\theta-2\xi/r})$. Eq.(\ref{iseff})
yields non-zero values of the integral $I$ {\it below} a cut-off
radius $r<(v_0t-l_0)/(1-{1\over 2}\sin^2\theta)$. However, the
asymptotic behavior of $I$ when $r$ is large is found by noticing
that $R_{min}\approx \xi/sin\theta$ and $R_{max}\rightarrow\infty$
in this limit. We then find
\begin{equation}\label{isefflim}
I\sim \exp\left(-{(r+l_0-v_0 t)^2\over 2\sin^2\theta D^2}\right)
\end{equation}
The essential point to retain is that the profile of the radial
outgoing pulse in the case $D>>l$ is a Gaussian whose dispersion
is of order $D$. Thus, the conclusion is that, in both cases $l>D$
or $D>l$, the outgoing wavefunction has always the form of a packet
with dispersion $\sigma_r$ of the same order as the {\it largest}
of the two quantum coherence lengths, i.e. $\sigma_r\sim\max(l,D)$.

Furthermore, we notice that in the case
$D>>l$ the dispersion $\sigma_r$ depends also on $\theta$.
A detailed investigation of the Bohmian trajectories in this case
is, however, not possible from a numerical point of view, because
the asymptotic formulae (\ref{iseff}) and (\ref{isefflim})
are not valid at distances $r<D$, where the scattering effects
take place. Thus in the sequel we limit ourselves to a detailed
investigation of the Bohmian trajectories in the case $l>>D$, while
a qualitative discussion of the case $D>>l$ will be made in
section 4, referring to the issue of the particles' arrival time
distribution.

\section{Quantum trajectories}
We now discuss the main features of the de Broglie - Bohm quantum
trajectories focusing on the case $l>>D$.

\subsection{Separator and quantum vortices}
The form of the trajectories can be found by carefully
examining the structure of the quantum currents $\mathbf{j}=(\hbar/2mi)
(\psi^*\nabla\psi-\psi\nabla\psi^*)$. The main remark is that, due
to Eqs.(\ref{psiin}) and (\ref{psioutl2}), the ingoing wavefunction
term (which has a Gaussian form both in the $R$ and $z$ directions)
has a falling exponential profile at large distances from the center
of the Gaussian, while the outgoing wavefunction has a more complex
form falling asymptotically as a power-law $1/r$ due to the factor
$f$. Thus, there is an inner domain of the quantum
flow where $\psi_{ingoing}$ prevails, and an outer domain where
$\psi_{outgoing}$ prevails. We call {\it separator} the boundary
delimiting the two domains. Formally, the separator is defined
as the (time-evolving) geometric locus where
\begin{equation}\label{separcon}
|\psi_{ingoing}| =|\psi_{outgoing}|
\end{equation}

In the case where the outgoing wavefunction is given by
Eq.(\ref{psioutl2}), the condition (\ref{separcon}) takes the form:
\begin{equation}\label{septimel}
\exp\left(-{R^2\over 2D^2}
-{(z+l_0-v_0t)^2\over 2l^2}\right)=
{|Z_1Z|q_e^2m\over 4\pi\epsilon_0\hbar^2}
{D\over a}\sqrt{d\over a}f(r,\theta)
\exp\left(-{(r+l_0-v_0t)^2\over 2l^2}\right)
\end{equation}
where we use the approximations $D^2+i \hbar t/m\simeq
D^2$ and $l^2+i \hbar t/m\simeq l^2$. We note that these
approximations hold within a range of parameter values relevant
to concrete experimental setups. For example, assuming that
incident particles have velocities of order $v_0\sim 10^8$m/s,
the time required to travel a distance of order $10^{-2}$ --
$10^{-1}$m (which is the typical size of an experimental
setup) is of the order of $t=10^{-10}$ -- $10^{-9}$s. On the
other hand, the typical coherence lengths in experiments e.g.
for electrons are of order 1$\mu$m or larger. Hence, $\hbar t/m$
is much smaller than $D^2$ or $l^2$.
\begin{figure}
\centering
\includegraphics[scale=0.65]{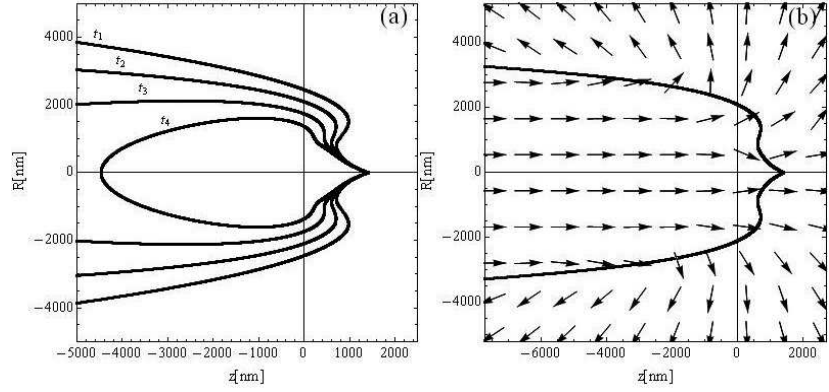}
\caption{(a) The form of the separator at four different time
snapshots $t_1=0$, $t_2=3l_0/(5v_0)$, $t_3=6l_0/(5v_0)$ and
$t_4=9l_0/(5v_0)$ in the model where $\psi_{ingoing}$ is given
by Eq.(\ref{psiin}), $\psi_{outgoing}$ is given by Eq.(\ref{psioutl2}),
and the parameters are $Z_1=-1$, $m=m_e$, $k_0=8.877\times
10^2$nm$^{-1}$ (corresponding to electrons with energy $E=30$KeV, or
wavelength $\lambda_0=7\times 10^{-3}$nm), $D=1000$nm, $l=10000$nm
(corresponding to transverse and longitudinal quantum coherence
lengths 1$\mu$m and 10$\mu$m respectively), $l_0=3l$, $Z=79$ (gold),
$d=420$nm, $a=0.257$nm. (b) The form of the quantum current
flow at the snapshot $t=t_2$. }
\label{septor}
\end{figure}
\begin{figure}
\centering
\includegraphics[scale=0.45]{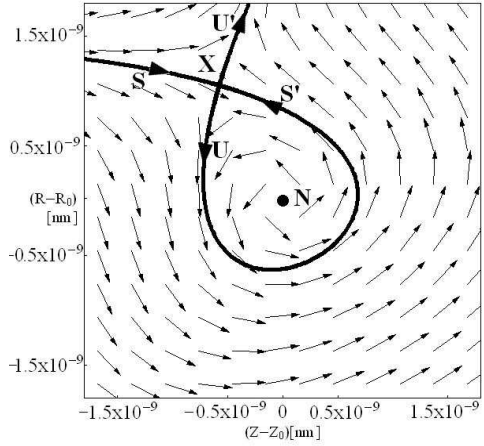}
\caption{Local form of the quantum flow at the `nodal point - X-point
complex' (quantum vortex) around the nodal point (N) with coordinates
$R=1934.42$nm, $z=137.178$nm in the model with parameters as in
Fig.\ref{septor} at the time $t=l_0/v_0$. The thick solid curves
show the unstable (U,U') and stable (S,S') asymptotic manifolds
of the X-point (X) formed under the instantaneous portrait of the
quantum flow.
}
\label{vortex}
\end{figure}

The time evolution of the separator in the plane $(R,z)$ depends
now on the time evolution of the relative amplitude of the ingoing
compared to the outgoing wave at any point of the configuration
space. We note first that according to Eq.(\ref{psioutl2}), the outgoing
wave corresponds to a wavepacket with dispersion $l$ which emerges
from the center in the time interval $t_0<t<t_0'$, with
$t_0=(l_0-l)/v_0$, $t_0'=(l_0+l)/v_0$, which is the interval during
which the support of the ingoing wavepacket (moving from left to right
in Figs.\ref{setup} and \ref{septor}) essentially overlaps with the
spatial domain occupied by the atoms in the target (see \cite{mes1961}
for an introductory description of this phenomenon in a simple
Rutherford scattering case). As indicated by Eq.(\ref{psioutl2}),
after its emergence the outgoing wave moves  in all radial directions
maintaining essentially its Gaussian profile, while its overall
amplitude drops like $r^{-1}$. As the outgoing wave moves
outwards, it first encounters the ingoing wavepacket at times
close to $t_0$. In Fig.\ref{septor}a, this encounter results in a
gradual approach of the separator towards the z-axis (the indicated
times are $t_1=0$, $t_2=3l_0/(5v_0)<t_0$, $t_0<t_3=6l_0/(5v_0)<t_0'$).
As, however, the ingoing packet moves from left to right in
Fig.\ref{setup}, its center crosses the target at the time
$t=l_0/v_0$. Afterwards, the ingoing wave emerges from the
right side of the target, and its support lies nearly completely
in the semi-plane $z>0$. At a still longer time ($t_4=9l_0/(5v_0)$),
the center of the outgoing wavepacket has traveled a distance
$\approx 2.5l$ apart, and there is no longer any overlapping
between the ingoing and outgoing wavepackets. As observed in
Fig.\ref{septor}a, a transition takes place at some time between
$t_3$ and $t_4$, such that, before the transition, the separator
is formed by a a pair of open curves on either side of the axis
$R=0$, while after the transition there is only one closed curve
intersecting twice the axis $R=0$ both for $z>0$ and $z<0$.
Taking into account the cylindrical symmetry around the z-axis,
the form of the separator in space before the transition is a
cylindrical-like surface of rotation, while after the transition
it becomes a prolate spheroidal-like surface. In fact, this
time-changing surface marks a sharp limit between the domains of
prevalence of the axial ingoing flow and the radial outgoing flow,
as shown in Fig.\ref{septor}b for the time $t=t_2$.

If Eq.(\ref{septimel}) is supplemented by an equation for the
phases of the ingoing and outgoing waves, which, for $Z_1<0$
takes the form
\begin{equation}\label{phaseop}
k_0R\tan(\theta/2)-\pi=2\bar{q}\pi~~~~~~~~\bar{q}\in{\cal Z}~~,
\end{equation}
a simultaneous solution of Eqs.(\ref{septimel}) and (\ref{phaseop})
defines the set of all points of the configuration space where the
total wavefunction (Eq.(\ref{psiinout})) becomes equal to zero.
Such points are called `nodal points'.

Around the nodal points, the quantum flow forms {\it quantum vortices}
(Figure \ref{vortex}). The local form of the quantum currents in a
vortex domain is very different from the general flow shown in
Fig.\ref{septor}. If we `freeze' the time $t$,
the instantaneous pattern formed by the vector field of quantum
probability current $\mathbf{j}$ corresponds to a characteristic
structure called {\it quantum vortex}, or {\it nodal point - X-point complex}
\cite{eftetal2007}\cite{coneft2008}\cite{eftetal2009}. That is, close to
a nodal point we find a second critical point of the flow, where one has
$\mathbf{j}=0$. This is called an `X-point', since it can be shown
that it is always simply unstable, i.e. there are two real
eigenvalues of the matrix of the linearized flow around X, which are
one positive and one negative. Accordingly, there are two opposite
branches of unstable (U,U') and stable (S,S') manifolds emanating
from X. On the other hand, the nodal point can be an attractor, center,
or repellor. This determines the local form of the invariant manifolds
U and S. It has been established theoretically \cite{eftetal2007}
that, except for a set of very small measure, most quantum
trajectories {\it avoid} the nodal point, being instead scattered
along the asymptotic directions of the manifolds of the X-point,
leading to large distances from the nodal point - X-point complex.
Furthermore, while, in general, the motion of the nodal point - X-point
complexes introduces chaos (\cite{fri1997}\cite{wispuj2005}
\cite{eftetal2007}\cite{eftetal2009}), in the present problem this
effect is negligible because i) the speed of vortices is extremely small
(of order $\sim\hbar/(k_0mD^2)<<v_0$), and ii) the quantum trajectories
exhibit only a small number of encounters with nodal point -
X-point complexes, as will be shown with numerical examples below.
In conclusion, the effect of the nodal point - X-point complexes
on the trajectories can be described as a scattering process
without recurrences.

For the model parameters used in Fig.\ref{vortex}, the size of the
quantum vortex, estimated by the distance $R_X$ from the nodal point
to the X-point, is of the order of $10^{-18}$m. The size of vortices
in the present model is in fact time dependent. However, in the time
interval $t_2\leq t\leq t_3$ when there is essential overlapping
of the ingoing and outgoing wavefunction terms, $R_X$ is approximately
constant and it is given by the same estimate as in the second of
the equations (25) of ref.\cite{deletal2011}, namely
\begin{equation}\label{rxest}
R_X=O\left({1\over Dk_0^2}\right)~~.
\end{equation}
The above estimate is obtained by expanding the wavefunction around
a nodal point up to terms of second degree in $(R-R_0)$ and $(z-z_0)$,
where $(R_0,z_0)$ are the coordinates of the nodal point, and by
applying general formulae derived in \cite{eftetal2009} regarding
the dependence of $R_X$ on the coefficients of this local expansion.

\subsection{Trajectories}
\begin{figure}[hpt]
\centering
\includegraphics[scale=0.75]{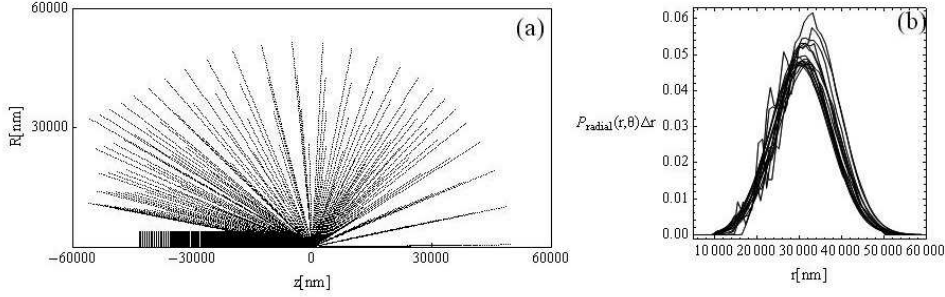}
\caption{(a) A swarm of Bohmian trajectories in the same
wavefunction model as in Fig.\ref{septor}. (b) The resulting
radial distribution $P_{radial}(r;\theta_i)$ for sixteen different
angles $\theta_i=5^\circ+(i-1)10^\circ$, $i=1,2,\ldots,16$.
The near coincidence of all distributions
$P_{radial}(r;\theta_i)$ and with the theoretical profile
corresponding to the outgoing wavefunction model (\ref{psioutl2})
indicates the degree of preservation of the continuity equation
by the numerical trajectories of (a).}
\label{orbitsl}
\end{figure}
The deflection of an orbit happens at the crossing of the separator,
and it is due to the orbit necessarily following the flow imposed by
the asymptotic manifolds of the X-points that exist along the separator.
Figure \ref{orbitsl}a shows a swarm of $625$ quantum trajectories in
the model with same parameters as in Fig.\ref{septor}. The initial
conditions are taken on a regular grid $25\times 25$ with
$-l_0-2l\leq z\leq -l_0+2l$ (where $l_0=3l$) and $D/100\leq z\leq 4D$.

An important test of the correctness of the numerical calculations
is by checking whether the probabilities associated with the chosen
initial conditions of the quantum trajectories respect the continuity
equation. To this end, setting $\psi\simeq\psi_{ingoing}$ at $t=0$,
a volume of initial conditions $\Delta V_0=2\pi R_0 \Delta R_0 \Delta
z_0$ centered around the point $(z_0, R_0)$ has an associated probability
$\Delta P= \mid \psi_{ingoing}(z_0,R_0,t=0)\mid^2 \Delta V_0$. Let
$(z_0,R_0)\rightarrow (r,\theta)$ be the mapping from initial
conditions to a trajectory' s coordinates at $t=2l_0/v_0$.
We want to estimate the mapping of the probabilities $\Delta P$
from the volume $V_0$ to the image of this volume under the
mapping $(z_0,R_0)\rightarrow (r,\theta)$. This is obtained
numerically, by quadratically interpolating first the functions
$r(z_0,R_0)$ and $\theta(z_0,R_0)$ from the data available
by the integration of the orbits with initial conditions on
the grid of points described in the previous paragraph. The
quadratic interpolation allows to obtain local approximations
to the functions $r(z_0,R_0)$ and $\theta(z_0,R_0)$ by formulae
of the form $r=A_0+A_1(z-z_0)+A_2(R-R_0)+A_3(z-z_0)(R-R_0)$,
$\theta=B_0+B_1(z-z_0)+B_2(R-R_0)+B_3(z-z_0)(R-R_0)$, where
the coefficients $A_i, B_i$ change values at every grid point.
These formulae, in turn, allow to numerically compute the
Jacobian determinant $J(r,\theta;z_0,R_0)= \Delta r
\Delta \theta /\Delta R_0\Delta z_0$. Finally, we compute
the probability function $P_{radial}(r,\theta)=
{\cal N}(\theta)|\psi_{in}(z_0,R_0,t=0)|^2R_0J(r,\theta;z_0,R_0)$,
where $z_0,R_0$ are functions of $(r,\theta)$ and ${\cal N}$ is a
normalization constant.

Figure \ref{orbitsl}b shows $P_{radial}$ as function of $r$ for
16 different values of $\theta$ in the interval
$5^\circ\leq\theta\leq 165^\circ$. The fact that all curves
nearly coincide implies that the numerically computed Bohmian
trajectories respect the continuity equation of the quantum
flow.

\begin{figure}[hpt]
\centering
\includegraphics[scale=0.65]{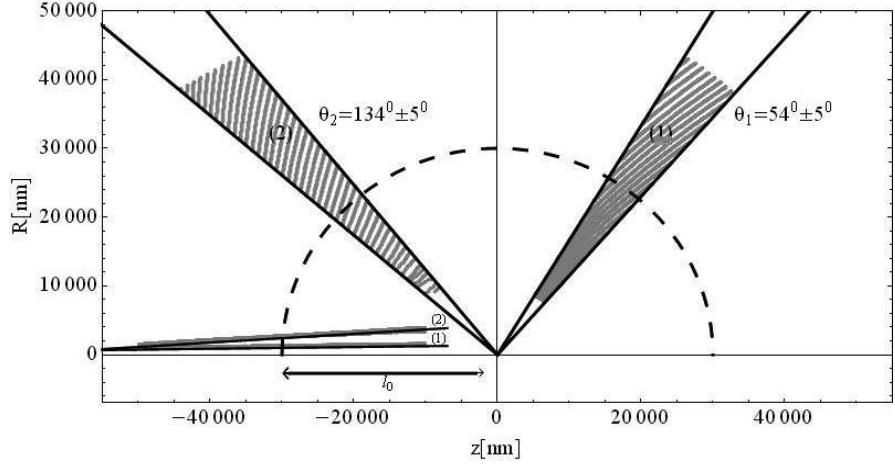}
\caption{The nearly straight gray zones in the bottom left
side correspond to the loci (1 or 2) of initial conditions
for which the corresponding de Broglie - Bohm trajectories
end in the angular sectors (1) $\theta=54^\circ\pm 5^\circ$
and (2) $\theta=134^\circ\pm 5^\circ$ at the end of the
numerical integration. The black lines on top of the gray
zones (1) and (2) correspond to the fitting by Eq.(\ref{inil3}).}
\label{zonelong}
\end{figure}
In the analysis of the arrival times or the times of flight in
section 4, use is made of the following information: we seek to
determine, as a function of the scattering angle $\theta$,
the locus of all initial conditions on the $(z_0,R_0)$ plane,
whereby the trajectories are eventually scattered close to the
angle $\theta$.

The determination of these loci follows by approximating all the
quantum trajectories as {\it piecewise straight lines}. Namely,
from Fig.\ref{orbitsl}, it is evident that any trajectory can
be considered as a nearly perfect horizontal line up to the
point (in space and time) where the trajectory encounters the
separator. In fact, as we can see in Fig.\ref{septor}, the
general motion of the separator itself, as $t$ increases,
is downwards. That is, if we consider a ray from the center
outwards along any fixed value of the angle $\theta$, the
separator intersects this ray at a continually decreasing
value of $r$, denoted by $r_s(t;\theta)$. Since all trajectories
are horizontal before the encounter, we have that $R(t)=R_{(0)}$ for
the z-coordinate of a trajectory with initial conditions
$(z_{(0)},R_{(0)})$. Then, the encounter takes place at the time
$t=t_{coll}$ when $R(t_{coll})=R_{(0)}=r_s(t_{coll};\theta)\sin\theta$.
The last condition determines the time $t_{coll}$, which is given by
\begin{equation}\label{tcoll1}
t_{coll}={-z_{(0)}+R_{(0)}\cot\theta\over v_0}
\end{equation}
Substituting Eq.(\ref{tcoll1}) in the separator equation
(\ref{septimel}), with $R=R_{(0)}$ we find:
\begin{equation}\label{inil1}
-{l^2g(\theta)\over 4D^2}R_{(0)}+{1\over g(\theta)}R_{(0)}
+z_{(0)}+l_0=
{l^2g(\theta)\over 2R_{(0)}}
\ln\left({|C S_{eff}| g(\theta)\over 2k_0^2R_{(0)}}\right)
\end{equation}
where $g(\theta)={2\sin \theta/{(1-\cos\theta)}}$ and
$C=mZ_1 Zq_e^2/(4\pi\epsilon_0 \hbar^2)$.
The last equation allows to determine $R_{(0)}$ as a function
of $z_{(0)}$. In the limit $l^2g^2(\theta)/(4D^2)>>1$,
we find an approximative formula by replacing the r.h.s.
of Eq.(\ref{inil1}) by a constant average value, i.e.
\begin{equation}\label{inil3}
R_{(0)}=R_c+{4D^2(z_{(0)}+l_0)\over l^2g(\theta)}
\end{equation}
where we take $R_c$ equal to the root for $R_{(0)}$ of
Eq.(\ref{inil1}) when $z_{(0)}=z_c=-l_0$. Eq.(\ref{inil3})
is an analytical expression which gives the locus of initial
conditions of trajectories that are scattered close to the
angle $\theta$. Figure \ref{zonelong} shows
lines of this form for two angles $\theta_1=54^{0}$,
$\theta_2=134^{0}$ along with the loci, at $(t=2l_0/v_0)$,
formed by the final points of the trajectories scattered
in a bin around these two angles.

\begin{figure}
\centering
\includegraphics[scale=0.7]{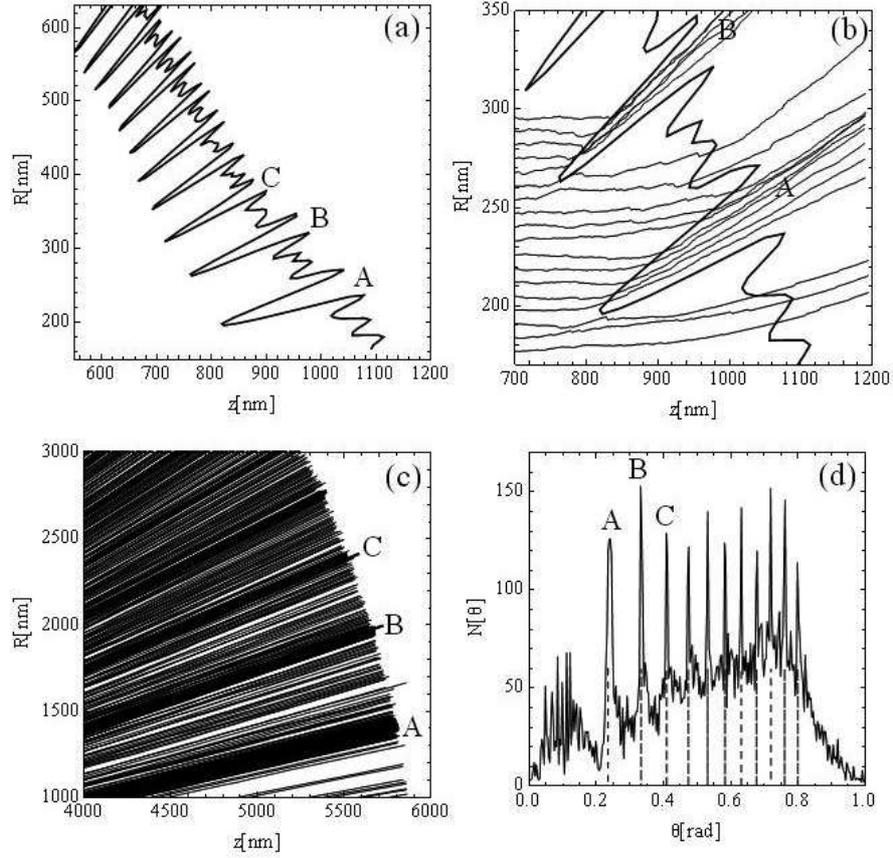}
\caption{(a) Local deformation of the separator at the time $t=l_0/v_0$
after the inclusion of the Bragg angles $\theta_q$ (Eq.\ref{bragg}) in the
effective Fraunhofer function (we mark the first three angles as
$A,B,C$). (b) The deflection of the Bohmian trajectories
at the channels of radial flow formed around the Bragg angles
A,B corresponding to $\theta_1=0.23...$ and $\theta_2=0.33...$
respectively). (c) The concentration of the scattered
trajectories close to Bragg angles in a larger scale. (d) Angular
distribution corresponding to the numerical trajectories of (c).
The dashed lines denote the exact positions of the Bragg angles.
}
\label{diforb}
\end{figure}

\subsection{Emergence of the diffraction pattern}
As mentioned in subsection 2.1, Eq.(\ref{psioutl2}) provides an
approximation to the outgoing wavefunction for nearly all sets
of values $(k_0,\theta,\phi)$ except very close to combinations
resulting in the appearance of a diffraction pattern. For specific
values of $k_0$ this pattern can be non-axisymmetric. Here, however,
we examine for simplicity only the appearance of Bragg angles for
which the resulting diffraction pattern is axisymmetric. This
implies considering the double sum over $n_x$, $n_y$ in
Eq.(\ref{sefflong}) as a sum of random phasors, while allowing
for a coherent addition of the phasors in the second sum of
(\ref{sefflong}). Eq.(\ref{seff}) takes the form
\begin{equation}\label{seffdif}
S_{eff}(k_0,\mathbf{r},t)\simeq
(D/a)e^{-{(r+l_0-v_0t)^2\over 2(l^2+{i\hbar t\over m})}}
\sum_{n_z=-N_z/2}^{N_z/2}
e^{ik_0[(1-\cos\theta)n_za+n_z^2a^2/(2r)]}~~.
\end{equation}
In order to estimate the sum in the r.h.s. of (\ref{seffdif}),
we first note that coherent contributions come only from atoms
whose z-position satisfies the condition $k_0z_j^2/(2r)<1$.
The coherent terms appear at the Bragg angles
\begin{equation}\label{bragg}
\sin^2(\theta_q/2)={q\pi\over k_0a},~~~q=1,2,...q_{max}
\end{equation}
Expanding the terms in the phase of $S_{eff}$ depending on $\theta$
around one Bragg angle we find
\begin{equation}\label{seffbragg}
S_{eff}(k_0,\mathbf{r},t)\sim(D/a)\sum_{n_z=-n_{z_0}}^{n_{z_0}}
e^{ik_0[\sin\theta_q(\theta-\theta_q)n_za
+{n_z^2a^2\over 2r}]}
\end{equation}
where $n_{z_0}\sim[(1/a){({r\over k_0})^{1/2}}]$. Exploiting the foil's
symmetry in the $z$ direction we approximate the sum in (\ref{seffbragg})
as
\begin{equation}\label{sall}
S_{eff}(k_0,\mathbf{r},t)\sim{ 2(D/a)(1/a)[\int_{0}^{u_{max}}du
e^{ik_0u^2\over{2r}}-{1\over 2}\int_{0}^{u_{max}}}du
e^{ik_0u^2\over{2r}}k_0^2\sin^2\theta_q(\theta-\theta_q)^2u^2]
\end{equation}
where $u_{max}\sim(r/k_0)^{1/2}$. An explicit formula for the
above integral can be given in terms of error functions. However,
a qualitative understanding of its behavior is offered by the
approximation $e^{ik_0u^2\over{2r}}\simeq
{1+{ik_0u^2\over{2r}}}$, whereby it follows that
$ \int_{0}^{u_{max}}du e^{ik_0u^2\over{2r}}\simeq
{e^{ik_0u_{max}^2\over{2r}}(u_{max}-{{1\over{3}}{ik_0\over
r}u_{max}^3})}$ and $\int_{0}^{u_{max}}du
e^{ik_0u^2\over{2r}}u_{max}^2=
{{1\over{3}}u_{max}^3}e^{ik_0u_{max}^2\over{2r}}$.
Substituting the above expressions in (\ref{sall}) we find
\begin{equation}\label{seffbragg2}
S_{eff}(k_0,\mathbf{r},t)\sim{ 2(D/a)(1/a)
e^{ik_0u_{max}^2\over{2r}}[u_{max}-{1\over3}
{{ik_0\over r}}u_{max}^3-{1\over
6}k_0^2\sin^2\theta_q(\theta-\theta_q)^2u_{max}^3]}
\end{equation}
Taking into account also the diffuse term, the final form
of the outgoing wavefunction is
\begin{equation}\label{psioutfinal}
\psi_{outgoing}\simeq \ 2{B(t)Z_1Zq_e^2m\over
4\pi\epsilon_0\hbar^2}(D/a) e^{-{(r+l_0-v_0
t)^2\over{2(l^2+{i\hbar t\over m})}}} f(r,\theta) e^{ik_0r}
\left[\sqrt{d\over a}+\sum_q U_q(r,\theta)e^{i\Phi_q(r,\theta)}\right]
\end{equation}
where the sum is considered with respect to all Bragg
angles, while the following estimates hold for the functions
$U_q$ and $\Phi_q$:
\begin{equation}\label{uq}
U_q\sim \frac{2\sin\left[k_0r\sin(\theta_q)(\theta-\theta_q)/2\right]}
{k_0a\sin(\theta_q)(\theta-\theta_q)}\nonumber\\
\end{equation}
and
\begin{equation}\label{phiq}
\Phi_q\sim \tan^{-1}
\left({1\over{-3+{1\over 2}rk_0\sin^2\theta_q(\theta-\theta_q)^2}}\right)
\end{equation}
sufficiently far from the target. The last equation implies that
at angular distances $|\theta-\theta_q|\sim \pi/(rk_0)^{1/2}$,
the particles' Bohmian trajectories acquire a transverse velocity
$v_t=(1/r)\partial\Phi_q/\partial\theta$ pointing towards the
direction of the straight line with inclination equal to
$\tan\theta_q$, while $v_t=0$ exactly at $\theta=\theta_q$.
Furthermore, the presence of the coherent terms in $S_{eff}$
causes a local deformation of the separator around the Bragg
angles, as shown in Fig.\ref{diforb}a. We note that the separator
comes locally closer to the center at the directions corresponding
to the Bragg angles, since the magnitude of $\psi_{outgoing}$ is
locally enhanced due to the local peaks of the functions $U_q$.

The effect of this deformation on the Bohmian trajectories is
analogous to the one described in \cite{deletal2011}. Namely,
this deformation results in the formation of local {\it channels
of radial flow}, whereby the Bohmian trajectories are preferentially
scattered around the Bragg angles. An example of this concentration
is shown in Fig.\ref{diforb}b. Clearly, the inclusion of the
coherent terms causes a variation of the angular distribution
of the Bohmian trajectories, by creating local maxima of
the density around the Bragg angles ($\theta_1=0.23...$,
$\theta_2=0.33...$ in Fig.\ref{diforb}b). Figure \ref{diforb}c
shows this concentration in a larger scale, while Fig.\ref{diforb}d
shows the angular distribution corresponding to the trajectories of
Fig.\ref{diforb}c. This distribution exhibit clear peaks at all
the angles $\theta=\theta_q$ (the first local maximum around
$\theta=0.1$ is not due to a concentration at a Bragg angle,
but it is only caused by the trajectories moving nearly
horizontally, i.e. within the support of the ingoing
wavepacket). We note that plots of the quantum trajectories
in a different scattering problem (atom surface scattering),
appearing in \cite{sanzetal2004a}\cite{sanzetal2004b}, show
a similar qualitative picture as in Fig.\ref{diforb}b, a
fact which was identified in that case too as a dynamical
effect of the quantum vortices. We may thus conjecture that
the quantum vortices play an important role in a wide context
of different quantum-mechanical diffraction problems.

Finally, it should be stressed that the modification of the outgoing wavefunction
according to Eq.(\ref{psioutfinal}) only influences the Bohmian velocity
field in the transverse direction, while the radial flow of all Bohmian
trajectories (as in Fig.\ref{diforb}) takes place at a constant speed
$\hbar k_0/m$. Thus, the emergence of a diffraction pattern does not
influence estimates on times of arrival or the times of flight of the
particles to detectors placed at the same distance from the center,
independently of the angle $\theta$. This subject is now discussed
in section 4.

\section{Arrival times and times of flight}
An important practical utility of the quantum trajectory approach
regards the possibility to unambiguously determine the probability
distributions of the so-called {\it arrival times}, or of the
{\it times of flight} of the scattered particles.  This question
is of particular interest, because it is related to the well known
`problem of time' in quantum theory (see \cite{muglea2000,mugetal2002}
for reviews). This problem stems from a theorem of Pauli \cite{pau1926},
according to which it is not possible to properly define a self-adjoint
time operator consistent with all axioms of quantum mechanics. This
implies that the usual (Copenhagen) formalism based on state
vectors or density matrices is not applicable to a quantum-theoretical
calculation of probabilities related to time observables. In fact,
in both Schr\"{o}dinger's and Heisenberg's pictures, time is
considered only as a parameter of the quantum equations of motion.

Among various proposals in the literature aiming to remedy this gap
of standard quantum theory (\cite{muglea2000}), the Bohmian formalism
offers a straightforward solution. This is in principle subject to
experimental testing, as will be proposed below. Furthermore, as
was mentioned in the introduction, the use of a wavepacket approach
allows for a comparison of the Bohmian approach with two other main
approaches to the same subject, namely the `history approach'
(based on Feynman paths) \cite{hartle1988}\cite{yamtak1993},
and the Kijowski approach, based on so-called `Bohm-Aharonov
operators \cite{kij1974}. We should
emphasize, however, that so far in the literature the latter
approaches were given a consistent formulation only in the case
of asymptotically free wavepacket motion \cite{muglea2000},
while their implementation in the case of scattered wavepackets
is an open issue. This will be discussed in a future work.
In the present paper, on the other hand, we focus on results
regarding the time observables as defined in the Bohmian approach,
and only provide a rough comparison of what should be expected
by other theories of time observables.

Any definition of a time observable in quantum mechanics (and also
in classical physics) requires the occurrence of two events serving
as the `start' and the `stop' event in the process of timing.

In the definition of the {\it arrival times} of particles to detectors,
we take as start event the preparation of the whole initial
wavefunction at a certain moment $t=t_1$, which can be conveniently
set equal to $t_1=0$. The stop event is the detection of the particle
at a time $t=t_2>t_1$. The arrival time of the particle to the
detector is $t_{arrival}=t_2-t_1$.

The above definition of arrival times is independent of the adopted
picture of quantum mechanics, since its only requirement is to assume
that the preparation of an initial quantum state can be controllable
in time (see \cite{bal2008}), i.e. that replicas of the same state can
be prepared at any given time $t_1$ (a technique realizing this
experimentally will be proposed below). However, a consistent
calculation of the arrival time probabilities by the various
pictures is still an open theoretical issue (see \cite{muglea2000}).

On the other hand, the definition of a {\it time of flight} for
a particle depends on the adopted picture of quantum mechanics,
since it requires the use of some notion of {\it spacetime paths}
that the particles presumably follow within the picture's
framework. In fact, the time of flight is defined as the
time elapsing between the crossing by the particle of two
surfaces $S_1$ and $S_2$ in the configuration space. Thus,
this time is different from the arrival time, and the difference
depends on where exactly the particle lies within the support of
the initial wavefunction.

In the case of particle diffraction, it is convenient to choose
the two surfaces as shown in Fig.\ref{setup}: $S_1$ is taken
normal to the z-axis at a point $z=-l_0$, while $S_2$ is
a spherical surface surrounding the target at the radial
distance $r_2=l_0$. The time of flight $T(\theta)$ depends
on the scattering angle $\theta$, and the quantity of interest
is the difference $T(\theta_2)-T(\theta_1)$ for two angles
$\theta_1$, $\theta_2$. This difference is independent of
$l_0$, provided that $l_0$ is such that both $S_1$ and $S_2$
are sufficiently far from the target.

We now discuss separately the arrival time and the time-of-flight
probabilities in the setup of Fig.\ref{setup}.

\subsection{Arrival times}

The motion of all scattered particles (like in Fig.\ref{orbitsl})
in the domain beyond a sphere of radius $2l_0$ around the center
can be considered as a radially outward motion with constant speed,
since, by taking $\psi\simeq\psi_{outgoing}$ (where $\psi_{outgoing}$
is given by Eq.(\ref{psiout})), the Bohmian equations of motion in
spherical coordinates read $dr/dt=v_0=\hbar k_0/m$, $d\theta/dt=0$.
The last equation is modified when the diffraction terms in $S_{eff}$
are taken into account. This modification, however, does not influence
the motions in the radial direction, which, as shown now, are the
only ones affecting the distribution of arrival times to a detector
placed at a distance $l_D$ and at any fixed angle $\theta$.
By definition, the latter distribution is given by
\begin{equation}\label{arrtimedis}
P_{arrival}(t)={\Delta N_{\theta,l_D}(t)\over\Delta t}
\end{equation}
where $\Delta N_{\theta,l_D}(t)$ denotes the number of particles
within the (assumed fixed) detector conic aperture
$d\Omega_D=\sin\theta\Delta\theta_D\Delta\phi_D$ around the
angle $\theta$ arriving to the detector between the times $t$ and
$t+\Delta t$. Since in the Bohmian approach all the particles move
with constant speed, we have
$$
{\Delta N_{\theta,l_D}(t)\over\Delta t}
=
{\Delta N_{\theta,l_D}(t)\over\Delta r}
{\Delta r\over\Delta t}=l_D^2d\Omega_D\rho v_0=
l_D^2d\Omega_D|\psi_{out}|^2{\hbar k_0\over m}
$$
In the case $l>>D>>a$, substituting Eq.(\ref{psioutl2}) and making the
usual approximations $l^2>>\hbar t/m$, $D^2>>\hbar t/m$ results in
\begin{equation}\label{arrtimedisl}
P_{arrival}(t)\simeq
P_0 e^{-(l_D+l_0-v_0t)^2\over \ell^2}
\end{equation}
where $P_0$ is a normalization constant. On the other hand, in the
case $D>>l>>a$ (via Eq.(\ref{isefflim}) we find
\begin{equation}\label{arrtimediss}
P_{arrival}(t,\theta)=
P_0' e^{-(l_D+l_0-v_0t)^2\over \sin^2\theta D^2}
\end{equation}
The main result can be summarized as follows: in either case
$l>>D$ or $D>>l$, the arrival time distribution is a localized
distribution (Gaussian, around the mean time $(\ell_D+\ell_0)/v_0$),
whose dispersion is always of the order of {\it $v_0^{-1}$ $\times$
the maximum of the transverse and longitudinal coherence lengths}.

The latter property implies that the trajectory approach makes
predictions regarding the arrival time distribution which depend
on two main beam parameters, and are thus testable in principle
by concrete experimental setups. One possible proposal in this
direction is the use of the so-called {\it laser-induced cold
field emission technique} (see \cite{baretal2007}). In this
technique, a cold-field electron source (nanotip) is exposed
to well separated in time focused weak laser pulses of time
width $\sim 10$--$100$fs. The photo-emitted electrons are accelerated
towards the anode, whereby their initial state can be effectively
described by an ingoing wavefunction of the form (\ref{psiin}).
The scattered particles pass through detector placed at fixed
angles $\theta$ as indicated in Fig.\ref{setup}. The key point to
notice is that time measurements in such a setup conform with
the definition of the {\it arrival times}, since a detection
of the triggering laser pulse can serve as a start event
marking the initial time when the {\it whole state} $\psi_{in}$
was prepared, while a later detection of a scattered particle
serves as the stop time $t_2$. We propose that by monitoring the
electron beam one can achieve different values of $l$ and $D$,
thus probing quantitatively the predictions for $P_{arrival}(t,\theta)$
as given by the quantum trajectory approach.

\subsection{Times-of-flight}

\begin{figure}[hpt]
\centering
\includegraphics[scale=0.5]{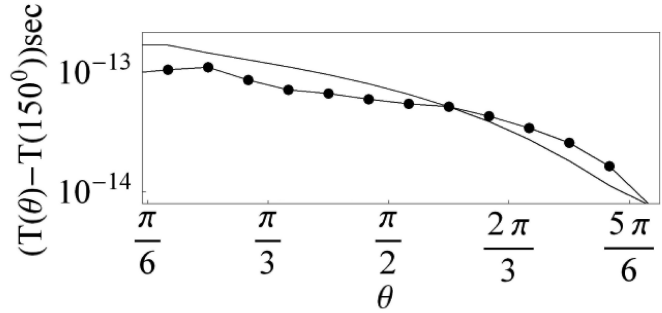}
\caption{The time difference $T(\theta)-T(150^\circ)$ (see text)
for the Bohmian trajectories of Fig.\ref{orbitsl}. The smooth
solid curve is the theoretical prediction of Eq.(\ref{dtth2})
while the dots represent numerical results.}
\label{timeavanglel}
\end{figure}

The total time of flight from a point on $S_1$ to a point on $S_2$
is a function of the initial conditions $(z_0,R_0)$. Using the
information from the swarm of numerical Bohmian trajectories of
Fig.\ref{orbitsl}, this function can also be quadratically interpolated
by the numerical data on grid points. The mean time $T(\theta)$
for all initial conditions leading to the same $\theta$ is then
found numerically. Since the choice of surfaces $S_1$ and $S_2$
in Fig.\ref{setup} is arbitrary, the invariant quantity of interest
is the difference $T(\theta)-T(\theta_0)$, where $\theta_0$ is a
fixed reference angle. Figure \ref{timeavanglel} shows this difference
for the numerical trajectories of Fig.\ref{orbitsl}a.

To estimate $T(\theta)$ theoretically, we make use of Eq.(\ref{inil3}),
yielding the locus ${\cal L}(\theta)$ of all initial conditions leading
to a scattering close to the angle $\theta$.  Using also the separator
equation (Eq.\ref{septimel}), we also find the point $(z_s,R_s)$
where the moving separator encounters an orbit moving horizontally
from $(z_{{\cal L}(\theta)}(R),R)$ at $t=0$, with speed $v_0$. Thus,
we set $R=R_s$ and $z_s=R_s/\tan(\theta)$. The time of flight of this
trajectory from $S_1$ to $S_2$ is then
$t(z_{{\cal L}(\theta)}(R),R)=(z_s+2l_0-l_1-R/\sin\theta)/v_0$.
The mean time of flight $T(\theta)$ can then be approximated
by $T(\theta)\approx \int_{{\cal L}(\theta)} 2\pi
R|\psi_{in}(z_{{\cal L}(\theta)}(R),R,t=0)|^2
t(z_{{\cal L}(\theta)}(R),R)dR$. We thus find:
\begin{eqnarray}\label{dtth2}
\Delta T=T(\theta_1)-T(\theta_2)&\approx &{DR_0\over v_0}
\left[\tan(\theta_2/2))- \tan(\theta_1/2)\right]~~~
\end{eqnarray}
where $R_0=[\sqrt{2\ln(C_{0})}+1/(1+\sqrt{2\ln(C_{0})})]$,
with $C_{0}=8\pi\epsilon_0k_0^2\hbar^2
/(|Z_1Z|q_e^2m\rho^{1/2}d^{1/2})$.

Due to Eq.(\ref{dtth2}), the time difference $T(\theta_1)-T(\theta_2)$ has an
$O(D\tan(\theta/2)/v_0)$ dependence on $\theta$. In fact, it is noticeable
that Eq.(\ref{dtth2}), which gives the difference of the mean times of flight
in the wavepacket approach, turns to be identical to the estimate of
\cite{deletal2011} (their Eq.(26)), referring to the plane wave approximation.
This is expected, since the plane wave approximation can be considered as
a limiting case of the wavepacket approach with $l>>D$, corresponding to
the limit $l\rightarrow\infty$.

One more interesting remark concerning the times of flight found by
the Bohmian approach is that the difference $T(\theta_1)-T(\theta_2)$
predicted in Eq.(\ref{dtth2}) has a completely different behavior from
analogous quantities calculated in the framework of other theories
of quantum time observables. While we defer a detailed reference to
this problem to a future work, here we give some rough estimates
concerning the sum-over-histories and the Kijowski approaches
referred to in the introduction. We find
\begin{eqnarray}\label{his}
\Delta T=
T(\theta_1)-T(\theta_2)\approx {-Z Z_1q_e^2\over 2\pi\epsilon_0m v_0^3}
\ln\left(\sqrt{1+\cot^2(\theta_1/2)\over 1+\cot^2(\theta_2/2)}\right)\\
~~~~~\mbox{(in sum-over-histories formalization,
semiclassical approximation)}\nonumber
\end{eqnarray}
and
\begin{equation}\label{kij}
\Delta T=T(\theta_1)-T(\theta_2)=0~~~~~\mbox{(in Kijowski formalization)}
\end{equation}
An outline of the derivation of these formulae is given in Appendix I.

A comparison of all three approaches yields that (i) the sum-over-histories
approach (which, using a semiclassical approximation yields essentially
the same result as in classical scattering theory), predicts a mean time
difference depending on the particle velocity $v_0$, by the scaling law
$\Delta T\sim [m_e/m][10^8\mbox{m sec}^{-1}/v_0]^3\cdot 10^{-19}sec$.
(ii) The Kijowski formalism predicts no time difference. (iii) The
Bohmian formalism predicts a time difference depending on $v_0$ as well
as on the transverse quantum coherence length $D$. The scaling
$\Delta T\sim [D/1\mu\mbox{m}] [10^8\mbox{m sec}^{-1}/v_0]10^{-13}$sec
holds.

As a final conclusion, we propose that experiments aiming to measure
time observables in setups of particle diffraction may provide
new insight into fundamental problems such as the role of time
in quantum mechanics. In particular, the predictions of the de Broglie
Bohm theory are within the possibilities of present day experimental
techniques.

\section{Semiclassical limit (Rutherford scattering)}
So far we have considered charged particles with quantum coherence
lengths much larger than the distance between nearest neighbors
in the target. However, this study does not cover the so-called short
wavelength limit, as e.g. in the case of $\alpha-$particle or
ion scattering. In this case, the quantum coherence length becomes
comparable to or smaller than the distance between nearest neighbors
in the target. As a result, such particles `see' each of the atoms
in the target as an individual scattering center and they do not
interact with the target lattice as a whole. Furthermore, both $D$
and $l$ become comparable to a classical `impact parameter' $b$
(of the order of a few fermi) which is relevant to the classical
description of Rutherford scattering.

The incorporation of $b$ in the wavefunction model can be done
essentially as described in \cite{mes1961}, assuming a
Gaussian form of his wavepacket, referred to by the notation $\chi$,
and aligning his vector denoted by $\mathbf{b}$ along the x-axis of
our coordinate system we have:
\begin{equation}
\psi(\mathbf{r},t)={1\over (2\pi)^{3/2}} \int
d^3\mathbf{k}~\tilde{c}(\mathbf{k})(1+f_k(\theta)){e^{ikr}\over
r})e^{{-i\hbar k^2t/2m}}
\end{equation}
with
\begin{equation}\label{psimom2}
\tilde{c}(\mathbf{k})={1\over \pi^{1/2}\sigma_\perp}
e^{(-{k_x^2+k_y^2\over 2\sigma_\perp^2})}{1\over
\pi^{1/4}\sigma_\parallel^{1/2}}
e^{{-(k_z-k_0)^2}\over{2\sigma_\parallel^2}}e^{-ik_xb}
\end{equation}
and
\begin{equation}\label{psimom2}
f_k(\theta)=-{Z_1Zq_e^2\over 4\pi\epsilon_0}{m\over
\hbar^2}{1\over{2\sin^2({\theta\over 2})k^2}}
\end{equation}
After the calculation of the above Gaussian integrals we are led to
the following wavefunction model:
\begin{equation}
\psi(\mathbf{r},t)=
\psi_{ingoing}(\mathbf{r},t)+\psi_{outgoing}(\mathbf{r},t)
\end{equation}
where
\begin{equation}\label{psiingse}
\psi_{ingoing}(\mathbf{r},t)= A\exp\left(-{(x-b)^2+y^2
+(z-v_0 t)^2\over 2(D^2+i\hbar t/m)} +i(k_0 z - \hbar k_0^2
t/2m)\right)
\end{equation}
\begin{eqnarray}\label{psioutse}
\psi_{outgoing}(\mathbf{r},t)&=& -A\left({Z_1 Zq_e^2\over
4\pi\epsilon_0}\right)
\left({m\over 2\hbar^2k_0^2\sin^2(\theta/2)r}\right)\\
&\times&\exp\left(-{(r-v_0 t)^2+b^2\over 2(D^2+i\hbar
t/m)} +i(k_0 r - \hbar k_0^2 t/2m)\right)\nonumber~~.
\end{eqnarray}
where
\begin{equation}
A={D\over{\pi^{1/2}}}({1\over{D^2+{{i\hbar
t}\over{m}}}}){\ell^{1/2}\over{\pi^{1/4}}}({l\over{\ell^2+{{i\hbar
t}\over{m}}}})^{1/2}
\end{equation}
is a nearly constant quantity, not affecting the Bohmian trajectories,
and $v_0=\hbar k_0/m$. The time $t=0$ in the above formulae is taken so that,
in the absence of scattering, the center of the ingoing wavepacket
crosses the plane $z=0$ at the moment $t=0$. Furthermore, we consider
the Bohmian trajectories at positive or negative times satisfying
$|t|<mD^2/\hbar$, i.e. smaller than the decoherence time of the packet.

The outgoing term is modulated by the Gaussian factor
$$
\exp\left(-{(r-v_0 t)^2+b^2\over 2(D^2+i\hbar
t/m)}\right)~~.
$$
This factor implies that a replica of the ingoing wavepacket
propagates from the center outwards as a spherical wavefront of the
outgoing wave, albeit by a phase difference $iv_0 t$ with
respect to the ingoing wavepacket. This new factor is the most
important for the analysis of Bohmian trajectories, because it
implies that the form of the latter depends crucially {\it on the
choice of the value of the parameter $b$}, which actually changes
the form of the wavefunction.
\begin{figure}[h]
\centering
\includegraphics[scale=0.8]{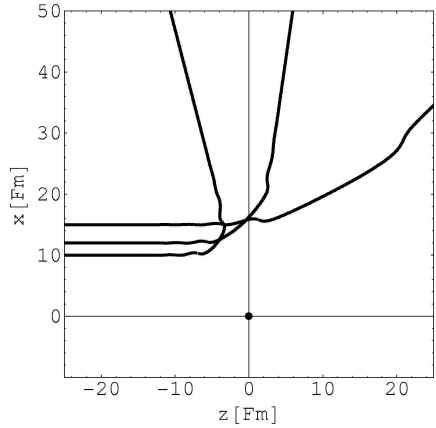}
\caption{Three bohmian trajectories guided by the wavefunction
defined in Eqs.(\ref{psiingse}) and (\ref{psioutse}), with
$D=l=10$fermi, $Z_1=2$, $m=7.1\times 10^3m_e$ (alpha particle),
$Z=79$ (gold), $k_0=10^{-14}m^{-1}$, and three different impact
parameters $b_1=10$fm, $b_2=12$fm, $b_3=15$fm. All three
trajectories are initially posed at the centers of their
corresponding guiding wavepackets.}
\label{ruth}
\end{figure}

A careful inspection of Eqs.(\ref{psiingse}) and (\ref{psioutse})
shows that the spherical wavefront emanating from $r=0$ encounters
the ingoing wavepacket at a time $t_c$ which {\it decreases as $b$
decreases}. As a result, the Bohmian trajectories, which are forced
to follow the motion of the radial wavefront after the collision,
are scattered to angles which are larger on the average for smaller
$b$. Thus, the Bohmian trajectories recover on the average the
behavior of the classical Rutherford trajectories. An example of
this behavior is given in Fig.\ref{ruth}, showing three Bohmian
trajectories corresponding to an initial condition taken at the
center of the wavepacket defined by Eq.(\ref{psiingse}) at the
time $t=-D^2m/\hbar$, ensuring that the packet's spreading does
not change appreciably in time up to the moment of collision with
the outgoing wavefront. The three Bohmian trajectories {\it cross
each other},  yielding a larger scattering angle for a smaller
initial distance from the z-axis, i.e. they are close to the
familiar classical picture. It should be noted, however, that
this closeness is only in an average sense, since the exact
form of a Bohmian trajectory guided by a wavepacket depends on
where exactly the initial condition of the trajectory lies with
respect to the center of mass of the initial packet. In fact,
for {\it one fixed value of b} one obtains a swarm of de
Broglie - Bohm trajectories (with initial conditions around
this value of $b$). These trajectories are scattered in various
directions and they do not cross each other, while they can
define (in a statistical sense) a most probable scattering angle.
However, this angle increases as $b$ decreases. Hence,
we conclude that when we consider the `semiclassical limit'
of small wavelengths as well as small quantum coherence lengths,
the Bohmian trajectories yield results which agree on the
average with the classical theory of Rutherford scattering.

\section{Conclusions}
We developed a wavefuntion model providing a wavepacket approach
to the phenomenon of charged particle diffraction from thin material
targets, and we employed the method of the de Broglie - Bohm
quantum trajectories in order to interpret the emergence of
diffraction patterns as well as to calculate arrival time
probabilities for scattered particles detected at various
scattering angles $\theta$. Our main conclusions are the
following:

1) In both cases when the longitudinal wavepacket coherence length
$l$ is larger than the transverse wavepacket coherence length $D$,
or vice versa, the outgoing wavepacket has the form of a pulse
propagating outwards in all possible radial directions, with a
dispersion $\sigma_r$ which is of the order of the maximum of $l$
and $D$. Furthermore, in the case $D>>l$ (applying e.g. to cold-field
emitted electrons), $\sigma_r$ depends on $\theta$ as $\sigma_r\sim
\sin\theta D$.

2) We study the structure of the quantum currents in the above model.
We provide theoretical estimates regarding the form and time evolution
of a locus called {\it separator}, i.e. the border between the domains
of prevalence of the ingoing and ourgoing quantum flow. We show how
the separator forms channels of radial flow close to every Bragg
angle, and leading to a concentration of the quantum trajectories
to particular directions giving rise to a diffraction pattern.

3) The deflection of quantum trajectories is due to their
interaction with an array of {\it quantum vortices} formed around a
large number of nodal points located on the separator. We show
examples of the quantum flow structure forming a `nodal point -
X-point complex' around any nodal point, and we calculate the
form of the stable and unstable manifolds yielding the local
directions of approach to or recession from an X-point. In
view of the similar role played by quantum vortices in different
examples of diffraction problems \cite{sanzetal2004a}
\cite{sanzetal2004b}, it can be anticipated that the mechanism
of emergence of the diffraction pattern described in section 3
is quite general.

4) We compute arrival time probability distributions for both
cases $l>>D$ and $D>>l$ using the de Broglie - Bohm trajectories
of particles detected at a fixed distance and various scattering
angles with respect to the target. In all cases, the dispersion
of the arrival time distribution turns to be $\sigma_t\sim
v_0^{-1}\times\max(l,D)$, where $v_0$ is the mean particles'
velocity. We propose a realistic experimental setup aiming to
test this prediction for electrons. We also calculate
time-of-flight differences, where the time of flight is defined
as the time interval separating the crossing by a de Broglie -
Bohm particle of two fixed surfaces located in the directions
of asymptotically free motion before and after the target.
We discuss the ambiguity of the definition of the times of
flight when using different approaches besides de Broglie
- Bohm, and provide a rough calculation in the framework
of the sum-over-histories approach and the Kijowski approach.

5) We finally examine how the de Broglie - Bohm trajectories
recover (in a statistical sense) the semiclassical limit of
Rutherford scattering, by examining the form of the quantum
trajectories when the packet mean wavelength, as well as $l$
and $D$ become smaller than the inter-atomic distance in the
target. In particular, we incorporate an impact parameter $b$
in the wavefunction model and demonstrate that the de Broglie
Bohm trajectories are scattered on the average at larger angles
$\theta$ as $b$ decreases. \\
\\
\noindent
{\bf Acknowledgments:} C. Delis was supported by the State Scholarship
Foundation of Greece (IKY) and by the Hellenic Center of Metals
Research. C.E. has worked in the framework of the COST Action
MP1006 - Fundamental Problems of Quantum Physics. He also acknowledges 
suggestions by H. Batelaan regarding the possibility to use the 
laser-induced field emission technique for arrival time measurements 
in the quantum regime.


\appendix
\section{Time-of-Flight differences outside the Bohmian formalism}

We hereby outline the derivation of Eqs. (\ref{his}) and (\ref{kij})
estimating the time-of-flight differences $T(\theta_2)-T(\theta_1)$
by the sum-over-histories and the Kijowski approach respectively.

{\it i) Sum-over-histories approach.} Neglecting the question
of consistency (see \cite{muglea2000}), a rough
calculation of the difference $T(\theta_2)-T(\theta_1))$ in the
sum-over-histories approach can be made in the framework of the
method developed in \cite{yamtak1993}. Denoting by $\Omega$ a fixed
space-time volume with time support within the interval from two
fixed times $t_A$ to $t_B$, the probability of a particle crossing
$\Omega$ is
\begin{equation}
P(\Omega)=\int d^3\mathbf{r}_B{\Bigg |}\int d^3\mathbf{r}_A
\Phi(B;\Omega;A)\psi(\mathbf{r}_A,t_A){\Bigg |}^2
\end{equation}
where $A\equiv(\mathbf{r}_A,t_A),B\equiv(\mathbf{r}_B,t_B)$, while
$\Phi(B;\Omega;A)=\sum_{\gamma\in\{B\leftarrow\Omega\leftarrow
A\}}e^{(i/\hbar)S(\gamma)}$ is the sum over all Feynman paths
$\gamma$ with fixed ends $A$ and $B$ passing through $\Omega$. The
quantity $P(t_A,t_B,\mathbf{r}_B;\Omega)=$ $\mid\int d\mathbf{r}_A
\Phi(B;\Omega;A)\psi(\mathbf{r}_A,t_A)\mid^2d^3\mathbf{r}_B$ is identified
as the probability that an electron being anywhere in space at
$t=t_A$ reaches a volume $\mathbf{r}_B+d^3\mathbf{r}_B$ at $t=t_B$
by passing first through $\Omega$. Let $S_1$ be a fixed surface normal to
the beam's central axis (Fig.\ref{setup}) at $-l_0<z_{S_1}<0$. We choose
$\Omega$ by the conditions that $\mathbf{r}_\Omega$ belongs to an
area element $\Delta S_1$ on $S_1$, around the point $R=0,z_\Omega=z_{S_1}$,
while $t_0\leq t_\Omega\leq t_0+\Delta t$, where $t_0>t_A$.
Let now $S_2$ be a spherical surface or radius $l_0$ around $O$,
and $\mathbf{r}_B$ a point on $S_2$ in the plane of Fig.\ref{setup}.
Setting $d^3{\mathbf r}_B= \Delta S_2v_0\Delta t$, where $\Delta S_2$
is an area element on $S_2$ around $\mathbf{r}_B$, the mean time of
flight from $\Delta S_1$ to $\Delta S_2$ becomes a function of $\theta$
only, given by
$T(\theta)=P_0\int (t_B-t_0) P(t_A=0,t_B,\mathbf{r}_B(\theta),\Omega) dt_0$
where $P_0=\left(\int  P(t_A=0,t_B,\mathbf{r}_B(\theta),\Omega) dt_0\right)^{-1}$.
To estimate $P(t_A,t_B,\mathbf{r}_B;\Omega)$ we extend results of
Hartle \cite{hartle1988} and Yamada and Takagi \cite{yamtak1993} in our case.
For $\Phi(B;\Omega;A)$ we adopt a 3D extension of a formula proposed
in \cite{yamtak1993}:
\begin{eqnarray}\label{phiyana}
\Phi(B;\Omega;A) = \int d^3\mathbf{r'}\Bigg[ \int d^3\mathbf{r}
\Phi(\mathbf{r}_B,t_B;\mathbf{r'},t_0+\Delta t)\nonumber\\
\Phi(\mathbf{r'},t_0+\Delta t;\Omega;\mathbf{r},t_0)
\Phi(\mathbf{r},t_0;\mathbf{r}_A,t_A)\Bigg]
\end{eqnarray}
where $\Phi(\mathbf{r},t_0;\mathbf{r}_A,t_A)$ is approximated by a
free Feynman propagator, $
\Phi(\mathbf{r}_B,t_B;\mathbf{r'},t_0+\Delta t) = \int d^3\mathbf{k}
e^{-{i\hbar k^2(t_B-t_0+\Delta t)\over 2m}}
\phi_\mathbf{k}(\mathbf{r'}) \phi_\mathbf{k}(\mathbf{r}_B) $
(with $\phi_\mathbf{k}$ as in Eq.(\ref{phiall})), while
$\Phi(\mathbf{r'},t_0+\Delta t;\Omega;\mathbf{r},t_0)$,
is approximated by a 3D analog of Hartle's approach \cite{hartle1988}
\begin{eqnarray}\label{hartle}
& &\Phi(\mathbf{r'},t_0+\Delta t;\Omega;\mathbf{r},t_0)\approx\nonumber\\
&=&\int_{t_0}^{t_0+\Delta t} dt \Bigg[
\int_{\Delta S_1}d^2\mathbf{R}_{S_1}
\left(m\over 2\pi i\hbar (t_0+\Delta t-t)\right)^{3/2}\nonumber\\
&\times&\left({|\mathbf{r}-\mathbf{r}_{S_1}|\over t-t_0}\right)
e^{{im\over\hbar}\left[{|\mathbf{r'}-\mathbf{r}_{S_1}|^2\over 2(t_0+\Delta t-t)} +
{|\mathbf{r}_{S_1}-\mathbf{r}|^2\over 2(t-t_0)}\right]}\Bigg]
\end{eqnarray}
where $\mathbf{r}_{S_1}$ are points on $\Delta S_1$ and
$\mathbf{R}_{S_1}=\mathbf{r}_{S_1}-z_{S_1}\mathbf{\hat{e}_z}$.
An exact calculation of all
integrals is untractable. Through a stationary phase approximation,
however, one obtains for fixed $t_A<0$ and
$\psi(\mathbf{r},t_A)\simeq\psi_{in}$, that
$P(t_A,t_B,\mathbf{r}_B;\Omega(t_0))$ is peaked essentially around a
mean `classical' time of flight for a trajectory starting from the center
of the wavepacket $|\psi_{in}|$, scattered by an atom at O, and arriving
to a point on $\Delta S_2$. For two angles $\theta_1$,
$\theta_2$, the mean time of flight difference can be estimated as
\begin{eqnarray}
T(\theta_1)-T(\theta_2)\approx {|Z Z_1| e^2\over 2\pi\epsilon_0m v_0^3}
\ln\left(\sqrt{1+\cot^2(\theta_1/2)\over 1+\cot^2(\theta_2/2)}\right)
\end{eqnarray}
i.e. we find Eq.(\ref{his}).

{\it ii) Kijowski approach}. The Kijowski approach \cite{kij1974},
assuming one-dimensional wave-packet propagation along some
direction $z$, we set
\begin{equation}\label{kij1}
\Pi(T,z)=
\sum_{s=-1,1}\Bigg|\int_{0}^{s\infty}dk\left({\hbar k\over m}\right)^{1/2}
\tilde{c}(k) e^{-i\hbar k^2 T/2m + ikz}\Bigg|^2
\end{equation}
to be the probability that the particle arrives on a normal surface at a
point $z$ between times $T$ and $T+dT$ ($\tilde{c}(k)$ is the
wavefunction in momentum space). If $\tilde{c}(k)\propto
e^{-(k-k_0)^2/2\sigma_\parallel^2 -ikz_0}$ is a narrow packet
($k_0>>\sigma_\parallel$), neglecting exponentially small negative
component terms of $\tilde{c}(k)$, we find
\begin{equation}\label{kij2}
\Pi(T,z)={hk_0\over m}\bigg[1+O\left(\sigma_\parallel/k_0\right)\bigg]|\psi(z,T)|^2
\end{equation}
i.e. $\Pi(T,z)$ practically coincides with the flux function
$J(z,T)=(hk_0/m)|\psi(x,T)|^2$. We apply Kijowski's
formalism in the setup of Fig.\ref{setup} for the asymptotically free
wavepacket motions at times long before or after $t=l_0/v_0$.
The mean time-of-flight from $S_1$ to a point of fixed $\theta$
on $S_2$ can be written as $T(\theta)=<t_2>-<t_1>$, where $t_1,t_2$ are
the arrival times to $S_1$ and $S_2$. Applying (\ref{kij2}) we
find $<t_1>=l_1/v_0$, $<t_2>=2l_0/v_0$. Thus, $T(\theta)=(2l_0-l_1)/v_0$
independently of the scattering direction, i.e.
\begin{equation}
T(\theta_1)-T(\theta_2)=0
\end{equation}
i.e. we find Eq.(\ref{kij}).

\end{document}